\UseRawInputEncoding
\documentclass[prl,preprint,preprintnumbers,amsmath,amssymb]{revtex4}
\usepackage{graphics,epsfig,epstopdf,stmaryrd}
\usepackage{textcomp}
\usepackage{bm}
\newcommand{\etal}{\textit{et al}.}
\newcommand{\rmh}{\text{Re}(E_p)} 
\newcommand{\al}{\alpha_{\rm{liss}}}
\begin {document}
\title{A predictive inline model for nonlinear stimulated Raman scattering in a hohlraum plasma.}
\author{D. B\'enisti$^{1,2}$}
\email{didier.benisti@cea.fr} 
\author{O. Morice$^1$}
\author{C. Rousseaux$^1$}
\author{A. Debayle$^{1,2}$}
\author{P.E. Masson-Laborde$^{1,2}$}
\author{P. Loiseau$^{1,2}$}
\affiliation{ $^1$CEA, DAM, DIF F-91297 Arpajon, France\\$^2$ Universit\'e Paris-Saclay, CEA, LMCE, 91680 Bruy\`eres-le-Ch\^atel, France.}
\begin{abstract}
In this Letter, we introduce a new inline model for stimulated Raman scattering (SRS), which runs on our radiation hydrodynamics code TROLL. The modeling  follows from a simplified version of a rigorous theory for SRS, which we describe, and accounts  for nonlinear kinetic effects.  It also accounts for the SRS feedback on the plasma hydrodynamics. We dubbed it PIEM because it is a fully PredIctivE Model, no free parameter is to be adjusted \textit{a posteriori}~in order to match experimental results. PIEM predictions are compared against experimental measurements performed at the Ligne d'Int\'egration Laser. From these comparisons, we discuss PIEM ability to correctly catch the impact of nonlinear kinetic effects on SRS. 
\end{abstract}
\maketitle
\begin{figure}[!h]
\centerline{\includegraphics[width=8.6cm]{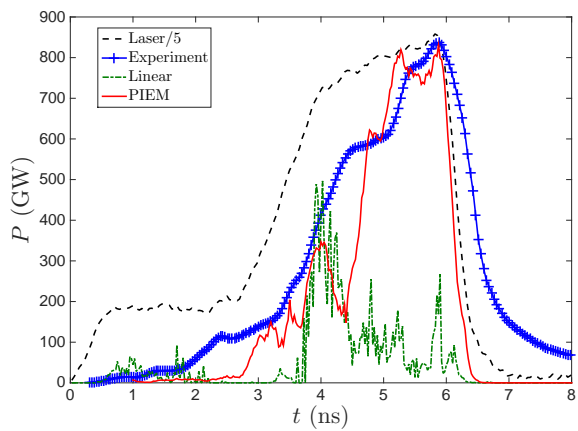}}
\caption{\label{retro} SRS reflected power as measured experimentally (blue curve with pluses), as predicted by PIEM (red solid line) and by a linear model (green dashed-dotted line). The black dashed line is the incident laser power divided by 5.}
\end{figure}

\textit{Introduction.}---When a large amplitude field acts on particles, it may nonlinearly modify their velocity distribution functions, with a potential feedback on the field itself. An effective modeling of these so-called nonlinear kinetic effects,  able to address large scale systems, has been a long-standing issue relevant to many fields of physics. These include, for example, electron acceleration by electrostatic waves in space plasmas~\cite{artemyev}, dark matter in cosmology~\cite{rampf}, or laser-plasma interaction and, in particular, stimulated Raman scattering (SRS)~\cite{kruer}. SRS has mainly been studied as a means to produce very large laser intensities by compressing energetic pulses~\cite{malkin}, and because it is detrimental to inertial confinement fusion (ICF). The latter effect, which  was the main motivation for our work, was clearly demonstrated  during the first experiments at the National Ignition Facilty (NIF)~\cite{meezan}. This was one motivation to introduce a new design, with low gas fill and shorter pulse duration, which successfully led to ignition~\cite{abu}. This new design, which effectively suppressed SRS, was found mostly empirically. There is still no way to make predictive simulations able to provide quantitative estimates for LPI, including Raman reflectivity, relevant to the ICF experiments at NIF or at the Laser M\'egajoule (LMJ). Yet, these estimates would be very useful for the introduction of new, and maybe more effective, designs. In order to accurately predict Raman reflectivity, one  would usually need a code that correctly estimates nonlinear kinetic effects. However, only a rad-hydro code can address the space and time scales relevant to ICF experiments. This calls for an inline model (i.e., which runs directly on a rad-hydro code), able to derive Raman reflectivity by accounting for nonlinear kinetic effects. This has long been thought as a very difficult, if not impossible, task. As a matter of fact, we are only aware of very few published inline models for SRS. Very recently, Stark~\etal~derived in Ref.~\cite{stark} a semi-empirical model based on a parameter study with two-dimensional particle-in-cell simulations, in essentially uniform plasmas. This model was designed to run on a rad-hydro code but has not been implemented in such a code yet. In Ref.~\cite{colaitis}, Cola\"{\i}tis~\etal~introduced a scaling law in the rad-hydro code CHIC in order to estimate the temperature of energetic electrons produced by SRS. In Ref.~\cite{strozzi}, Strozzi~\etal~introduced a linear kinetic modeling of SRS in the rad-hydro code Lasnex. However, this was a \textit{post hoc}~model, which required  the escaping SRS powers and wavelengths resulting from experimental measurements. In this Letter, we introduce the PIEM model which predicts the SRS reflected power and wavelengths without needing any input from experimental measurements. It has been implemented in our rad-hydro code TROLL~\cite{troll}~and allows for the SRS feedback on the plasma hydrodynamics. It is derived from a rigorous theory~\cite{ppcf,benisti20I,benisti20II} after several simplifying assumptions and,  as shown in Fig.~\ref{retro}, its predictions compare well with the experimental measurements reported in Ref.~\cite{lil}. However, PIEM predictions on SRS may only be accurate if all other LPI effects, and especially the plasma hydrodynamics, are also accurately computed in our rad-hydro simulation. When comparing our numerical results against experimental ones, we will be led to discuss TROLL's ability to correctly describe the plasma hydrodynamics. However, an in-depth discussion of the latter issue is way beyond the scope of this Letter. Here, we essentially discuss the hypotheses leading to our nonlinear kinetic modeling and, based on comparisons against experimental measurements, its ability to address such a large scale system as a hohlraum. \\
\textit{PIEM Model.}---PIEM essentially works as a nonlinear gain model, calculated along rays with curvilinear abscissa, $s$. Eqs.~(\ref{final1})~and~(\ref{final2}) express how the number of photons vary along a given ray, for the incident laser and backscattered waves. This depends on the electron plasma wave (EPW) amplitude $E_p$ which, from Eq.~(\ref{final3}), is calculated as a function of the product between the laser and backscattered waves amplitudes. Eq.~(\ref{final3}) actually is an implicit nonlinear equation for $E_p$, which makes one main difference between PIEM and a classical linear model. More precisely, PIEM equations for SRS are,
\begin{eqnarray}
\label{final1}
\partial_sA_l^2&=& -\frac{ek_p\rmh}{m\omega_b\sqrt{v_{g_l}v_{g_b}}}\sqrt{\frac{\zeta_b}{\zeta_l}} A_l[A_b+A_{b_f}],\\
\label{final2}
\partial_sA_b^2&=& -\frac{ek_p\rmh}{m\omega_l\sqrt{v_{g_l}v_{g_b}}}\sqrt{\frac{\zeta_l}{\zeta_b}} A_lA_b,\\
\label{final3}
E_p&=&\frac{\Gamma_p}{\gamma+\nu_p-i\delta\omega}\sqrt{\frac{v_{g_l}}{v_{g_b}}}\sqrt{\zeta_l\zeta_b}E_{l_0}^2A_l[A_b+A_{b_f}],
\end{eqnarray}
where the subscripts $_{l,b,p}$ are respectively for the laser, backscattered and plasma waves. $\zeta_{l,b}$ is the fraction of absorbed power by inverse bremsstrahlung, derived by accounting for the so-called Langdon effect, as described in Ref.~\cite{langdonIB}. For each wave, we use $\omega$ to denote the frequency, $k$ for the wavenumber, and $E$ for the amplitude. In Eqs.~(\ref{final1})~to~(\ref{final3}), $v_{g_{l,b}}=\vert k_{l,b}\vert c^2/\omega_{l,b}$ where $c$ is the speed of light in vacuum. In Eq.~(\ref{final3}), $E_{l_0}$ would be the laser wave amplitude absent of SRS and damping. Then, the laser and backscattered wave amplitudes, $E_l$ and $E_b$, are related to $A_l$ and $A_b$ by, $E_l=\sqrt{\zeta_l}A_lE_{l_0}$, $E_b=\sqrt{\zeta_bv_{g_b}/v_{g_l}}A_bE_{l_0}$. We also introduce $E_{b_f}$ as the amplitude of the fluctuations at frequencies close to $\omega_b$, which seed SRS. It is related to $A_{b_f}$ by $E_{b_f}=\sqrt{\zeta_bv_{g_b}/v_{g_l}}A_{b_f}E_{l_0}$. We did not try to make an accurate estimate for $E_{b_f}$ and we simply assumed $E_{b_f}=\eta E_{l_0}$. We checked that our results were essentially insensitive to our choice for $\eta$, and those shown in this Letter correspond to $\eta=10^{-5}$. Each set of equations (\ref{final1})-(\ref{final3}) is solved from $s=0$ to $s=s_{\max}$ which are, respectively, before the ray has entered and after it has exited the plasma. Then, the boundary conditions are  $A_l(s=0)=1$ and $A_b(s=s_{\max})=0$.  

 Note that, in  Eqs.~(\ref{final1}) and (\ref{final2}), we assume that the rays trajectories are identical for the laser and backscattered waves. We do not make this hypothesis  when accounting for the SRS feedback on the plasma hydrodynamics, as explained below, but only when solving the wave coupling equations. The hypothesis is clearly valid when the plasma is nearly uniform, in which case both rays are straight lines. When the plasma is inhomogeneous, SRS is only effective within the narrow space region where the resonance conditions, $k_l-k_b=k_p$ and $\omega_l-\omega_b=\omega_p$ are fulfilled. Within this region the laser and backscattered rays trajectories are nearly the same. TROLL derives these trajectories by making use of the geometrical optics approximation. Consequently, without SRS nor collisional damping, this code would predict an amplitude $E_{l_g}$ for the laser wave different from $E_{l_0}$. Indeed, in ICF experiments, the laser beams are optically smoothed~\cite{garnier}, which leads to an intensity pattern made of speckles. In order to approximately account for optical smoothing, we assume that $E_{l_0}^2(s)=\al(s)E_{l_g}^2(s)$, where $\al(s)\equiv\sum_i\alpha_i(s-\sigma_i)$. Here, $\alpha_i$ is a bell shaped function whose total width at half maximum is of the order of the longitudinal size, $l_{\parallel}$, of a speckle. We chose, $\alpha_i(s)=\alpha_{0_i}\sin_c^2[\pi (s/l_{\parallel})]$, where $l_{\parallel}=7\lambda_lf_{\#}^2$ ($f_{\#}$ being the overall aperture of the focusing system), and where $\alpha_{0_i}$ is an exponential random variable. Moreover, we impose the averaged value of $\al$  to be unity along each ray.  As for the $\sigma_i$'s, there are separated by $2l_{\parallel}$ and are time independent, so that we do not account for smoothing by spectral dispersion (SSD). This would impose too small timesteps in our rad-hydro simulation, while SRS usually grows too quickly to be directly sensitive to SSD. Our modeling for optical smoothing was designed to make PIEM as effective and simple as possible and we mainly rely on comparisons against experimental results, as illustrated in Fig.~\ref{retro}, to discuss its relevance. Finally, note that Eqs.~(\ref{final1}) and (\ref{final2}) are only valid when the waves amplitudes do not explicitly depend on time. Hence, we miss the distinction between a convective and an absolute linear instability~\cite{twiss,bers}. However, in this Letter, we address the nonlinear regime of SRS and we mainly want to estimate the maximum amplitude reached by the backscattered wave. In a rather homogeneous plasma, it follows from the limits imposed on the EPW amplitude by nonlinear saturation mechanisms, such as Langmuir Decay Instability (LDI)~\cite{ldi} and the growth of sidebands~\cite{friou}. In order to account for these, we assume $E_p<E_{\max}$ when solving Eq.~(\ref{final3}), where $E_{\max}$ is the minimum between the limit imposed by LDI~\cite{ldi} or by wavebreaking~\cite{benisti20II}.  In an inhomogeneous plasma, SRS is limited by the space extent of the region where the three-wave resonance is maintained, possibly with the help of autoresonance~\cite{autoresonance}. This is derived from Eq.~(\ref{final3}), from the magnitude of $\delta \omega$, which depends on the plasma inhomogeneity and the EPW amplitude thus allowing for autoresonance. More precisely, provided that $\gamma=E_p^{-1}[\partial_t+v_{g_p}\partial_s]E_p$ and $\delta\omega=(1+\chi)/\partial_\omega\chi$, Eq.~(\ref{final3}) is equivalent to the following envelope equation,
\begin{equation}
\label{e6}
\partial_{\omega}\chi\left[\partial_t+v_{g_p}\partial_s+\nu_p\right]E_p-i(1+\chi)E_p=\Gamma_p\partial_\omega\chi E_lE_b^{\rm{tot}},
\end{equation}
where we have denoted $E_b^{\rm{tot}}\equiv E_b+E_{b_f}$ and where $\Gamma_p\partial_\omega\chi=ek_p/m\omega_l\omega_b$, $-e$ being the electron charge and $m$ its mass. As for $\chi\equiv\chi(k_l-k_b,\omega_l-\omega_b,E_p)$, it reads
\begin{equation}
\label{e7}
\chi=(1-Y_{\rm{NL}})\chi_{\rm{lin}}+Y_{\rm{NL}}\chi_a,
\end{equation}
where $\chi_{\rm{lin}}$ is the linear electron susceptibility~\cite{linear}, $\chi_a$ is the adiabatic nonlinear susceptibility as derived in Ref.~\cite{benisti20II}, and $Y_{\rm{NL}}$ is the fraction of electrons which respond nonlinearly to the EPW. $Y_{\rm{NL}}$ has been derived in Ref.~\cite{ppcf} in a three-dimensional geometry and for a Maxwellian plasma, and reads
\begin{equation}
\label{e8}
Y_{\rm{NL}}= 1-\exp(-\omega_B^2 l_{\bot}^2/50v_{th}^2),
\end{equation}
where $v_{th}$ is the electron thermal speed, $\omega_B=\sqrt{2e\vert E_p\vert/m}$ is the so-called bounce frequency, and $l_{\bot}$ is an effective length scale for the transverse $E_p$ gradient. We chose $l_{\bot}=\lambda_lf_{\#}$. Eq.~(\ref{e8}) mainly expresses the fact that an electron responds nonlinearly to the EPW once it has completed one trapped oscillation within the wave trough. Then, $\nu_p=(1-Y_{\rm{NL}})\nu_L$, where $\nu_L$ is the Landau damping rate~\cite{landau}. Actually, Eq.~(\ref{e6}) corresponds to a simplified version of our theory, described for example in Ref.~\cite{benisti2007}. It neglects several effects such as the change in $E_p$ entailed by the convergence or divergence of rays. Moreover, in this equation, we use $v_g=-\partial_k\chi/\partial_\omega\chi$ which is not very accurate~\cite{dodin}. However, in an inhomogeneous plasma, the region where the resonance conditions are fulfilled is usually too narrow for the latter approximations to have a real impact on the EPW amplitude. Moreover, as shown in Ref.~\cite{benisti09,benisti2010,brama3D}, Eq.~(\ref{e6}) provides very accurate results for SRS in a homogeneous plasma. However, because we want PIEM to be an effective nonlinear gain model, we do not compute the actual value for $\gamma$. Instead, we use $\gamma=\sqrt{\gamma_0^2+\nu_p^2}-\nu_p$, where $\gamma_0=k_p v_{\rm{osc}}/\sqrt{2\omega_b\partial_\omega \chi_{lin}}$ with $v_{\rm{osc}}=eE_{l_0}/m\omega_l$. The latter approximate expression for $\gamma$ was successfully used in Ref.~\cite{dw} to account for the impact of the laser drive on the nonlinear frequency shift,  and in Ref.~\cite{benisti20II} to estimate the EPW wave front bowing. 

Along each ray, Eqs.~(\ref{final1})-(\ref{final3}) are solved for one single value of $\omega_b$. It is derived every $N_{w^*_b}$ timesteps as the frequency yielding the largest Raman reflectivity. For the simulation results presented below, we chose $N_{w^*_b}=15$. The chosen value for $\omega_b$ fulfills the three-wave resonance conditions at a given abscissa, $s^*$. Then, at $s^*$, we generate a backscattered wave that carries a number of photons, $N_b$, derived from the resolution of Eq.~(\ref{final2}). This wave propagates along its own ray 
 and deposits its energy in the plasma by inverse bremsstrahlung. Moreover, at $s=s^*$, the laser wave is depleted by $N_b$ photons. Hence, we allow for the SRS feedback on the plasma hydrodynamics.\\
 \textit{Numerical resolution}---Eq.~(\ref{final3}) for $E_p$ is solved by bisection~\cite{num_rec}, with $0\leq E_p\leq E_{\max}$, where $E_{\max}$ is the minimum between the limit imposed by LDI~\cite{ldi} or by wavebreaking~\cite{benisti20II}. In order to significantly speed up PIEM, we derive $\delta \omega$ from stored values obtained for a large range of electron densities and temperatures, and of EPW amplitudes. Eqs.~(\ref{final1}) and (\ref{final2}) are solved by making use of a fourth order Runge-Kutta method~\cite{num_rec}, with the space step $\delta s\approx 5$~$\mu$m. These equations are solved from $s=0$ to $s=s_{\max}$ for a given value of $A_b(0)$. Now, the value of $A_b(0)$ that yields our estimate for the reflected power is found by bisection. Knowing that $0\leq A_b(0)\leq\sqrt{\omega_b/\omega_l\zeta_b(0)}$ we solve Eqs.~(\ref{final1})-(\ref{final3}) until we find, using the bisection method, the value $A_b(0)$ that yields $A_b(s_{\max})\approx 0$. As for the plasma hydrodynamics, it was computed by making use of a two-dimensional axisymmetric TROLL simulation. The timestep used for the simulation was $\delta t=0.5$~ps. Note that, because PIEM calculates SRS along rays, its performance should not very sensitive to dimensionality.  In our simulation, we used $33^2=1089$ rays. \\
\textit{Comparisons against experimental results}---Let us now compare PIEM predictions against the experimental results detailed in Ref.~\cite{lil}, and obtained at the Ligne d'Int\'egration Laser (LIL). The laser system of the LIL facility consists of four square beamlets, put together into a quad~\cite{lill}, so that there cannot be any crossed beam energy transfer. Moreover, very little Brillouin reflectivity was measured during the experiments, so that nonlinear wave coupling seemed to be essentially limited to SRS. Hence, these experiments were particularly well suited to test the PIEM model. 
The total energy in the quad was close to 15.7~kJ and the temporal pulse shape, illustrated in Fig.~\ref{retro}, consisted of two plateaus, of about 3~ns each. The power in the first plateau was close to 1~TW (with a space-averaged intensity close to $2\times10^{14}$~W/cm$^2$), and the power in the second plateau was about  $4-4.5$~TW (which corresponds to a space-averaged intensity of about $8-9\times10^{14}$~W/cm$^2$). The quad was optically smoothed with kinoform phase plates and the pilot incorporated two phase modulators. The first one at 2 GHz prevented stimulated Brillouin scattering from growing within the optics while the second one, at 14 GHz, aimed at reducing LPI within the plasma. The quad was sent into a cylindrical hohlraum, 4 mm long and 1.4 mm diameter, filled with 1 atm neo-pentane gas. The hohlraum was closed by two thin polymide windows which exploded under the action of the quad. Hence, this was an open configuration. There were actually two similar experimental campaigns, one in 2011 whose results were published in Ref.~\cite{lil} and one in 2013. There was an excellent reproducibility as regards SRS reflectivity, and the results illustrated in Fig.~\ref{retro} were obtained in 2013. Experimentally, there is no way to exactly match the time origin of the reflectivity measurements with that of the incident pulse. A choice has to be made and, in this Letter, we chose to let the maxima of the  reflected and incident powers be at the same time. However, we can't exclude that they might be very slightly shifted, by at most 0.3~ns. As shown in Fig.~\ref{retro}, PIEM predictions for the SRS reflectivity are in good agreement with the experimental measurements. 
However, there are several discrepancies which we now explain. First, after 6~ns, the reflected power decreases more slowly in the experiment than in the simumation. This is most probably due to the slow response of the diagnostic. Indeed, when $t\agt6.5$~ns, the experimental reflectivity would be larger than unity. At $t\approx 4.5$~ns, the numerical prediction significantly underestimates the experimental result. This is most probably due to a poor estimate of the plasma hydrodynamics by TROLL. Indeed, when the laser quad enters the hohlraum, the plasma is first expelled from the propagation axis, and bounces against the hohlraum wall before coming back towards the axis when $t\approx 4$~ns. This is clear from the experimental reflected spectrum which exhibits two peaks at $t\approx 4$~ns (not shown here but illustrated in Ref.~\cite{lil}). Hence, there is plasma mixing, which TROLL usually does not address very well, and which most probably explains the discrepancy at $t\approx 4.5$~ns. Finally, before 3~ns, the simulation underestimates the measured reflectivity. As already reported in Ref.~\cite{lil}, before 3~ns, only a very small fraction of the incident laser power is transmitted outside the hohlraum, unlike what would predict the simulation and in spite of a very low reflectivity. This seems to indicate that the laser propagation or the plasma hydrodynamics are not well reproduced at early times by the TROLL simulation. Hence, at these times, it is difficult to draw any conclusion on PIEM accuracy from the comparisons against the experimental results. By contrast, comparisons against the predictions of the linear model described in Ref.~\cite{arnaud}, which has been recently implemented in TROLL, are very instructive. As shown in Fig.~\ref{retro}, this model systematically underestimates both, the PIEM  predictions and the experimental measurements, except when $t\approx4$~ns. Now, from our simulation results, we know where in the plasma most of the reflected light comes from. From the electron density and temperature at the corresponding position, we can estimate the value of $k_p\lambda_D$ ($\lambda_D=v_{th}/\omega_{pe}$ being the Debye length) and of the Landau damping rate, $\nu_L$, for the EPW resulting from SRS. 
\begin{figure}[!h]
\centerline{\includegraphics[width=8.6cm]{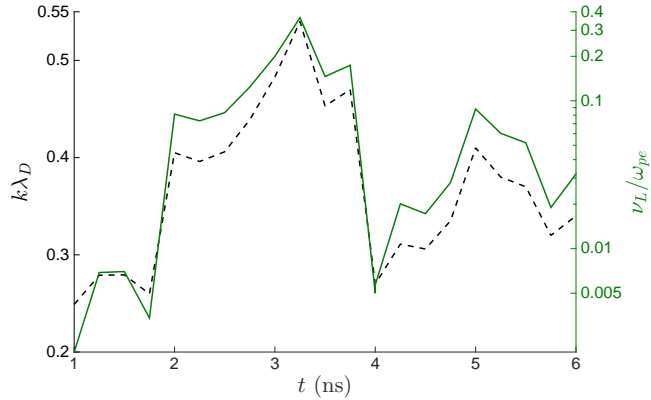}}
\caption{\label{kld} Time evolution of $k_p\lambda_D$ (black dashed line) and of $\nu_L/\omega_{pe}$ (green solid line) as calculated by PIEM.}
\end{figure}
As may be seen in Fig.~\ref{kld}, $\nu_L$ assumes a marked minimum when $t\approx 4$~ns.  It is several times smaller than $\gamma_0$, so that its nonlinear decrease has a very moderate impact on SRS. This is contrast with the situation at earlier and later times when $\nu_L$ assumes much larger values, respectively, because the density is smaller or because the temperature is larger. Actually, as shown in Fig.~\ref{retro}, $\nu_L$ grows so quickly after $t\approx 4$~ns that the linear reflectivity globally decreases although the laser power raises. This is in contrast with PIEM's prediction for the reflectivity, which globally increases after $t\approx 4$~ns, in agreement with the experimental results. This shows PIEM's ability to correctly account for the nonlinear reduction of the EPW damping rate, and more generally for the impact of nonlinear kinetic effects on SRS. Comparisons against linear results are a clear evidence of kinetic inflation~\cite{vu} and of PIEM's ability to allow for it. \\ 
 \textit{Conclusion}---In this Letter, we introduced the PIEM model, able to correctly predict the SRS reflected power measured on LIL experiments when nonlinear kinetic effects were clearly at play. This does not mean that we are able to make predictive ICF simulations for all possible situations. We do not claim that PIEM is perfect and complete. In the near future, we want to extend our theory for EPWs to ion waves in order to derive a nonlinear kinetic modeling for stimulated Brillouin scattering and crossed beam energy transfer that would be included in PIEM, in addition to SRS. We also want to allow for the production of hot electrons by SRS. Several issues, related to laser beam propagation, plasma hydrodynamics, or laser power absorption, may also need to be more accurately accounted for in TROLL. However, we do believe that PIEM opens the path to predictive simulations by successfully addressing one important issue, long thought as a particularly difficult one. It lets a radiation hydrodynamics code correctly estimate nonlinear kinetic effects. \\ \\
We thank C. Ruyer for his careful reading of the manuscript and his useful comments, and one of the authors (DB) acknowledges interesting discussions with D. Stark and L. Yin. 
 
\end{document}